# Microcavity design for low threshold polariton condensation with ultrashort optical pulse excitation

C. Poellmann,[1] U. Leierseder,[1] E. Galopin,[2] A. Lemaître,[2] A. Amo,[2] J. Bloch,[2] R. Huber,[1] and J.-M. Ménard[1,3,a]

[1]*Department of Physics, University of Regensburg, 93040 Regensburg, Germany*

[2]*CNRS-Laboratoire de Photonique et Nanostructures, Route de Nozay, 91460 Marcoussis, France*

[3]*Max Planck Institute for the Science of Light, Günther-Scharowsky-Straße 1, 91058 Erlangen, Germany*

We present a microcavity structure with a shifted photonic stop-band to enable efficient non-resonant injection of a polariton condensate with spectrally broad femtosecond pulses. The concept is demonstrated theoretically and confirmed experimentally for a planar GaAs/AlGaAs multilayer heterostructure pumped with ultrashort near-infrared pulses while photoluminescence is collected to monitor the optically injected polariton density. As the excitation wavelength is scanned, a regime of polariton condensation can be reached in our structure at a consistently lower fluence threshold than in a state-of-the-art conventional microcavity. Our microcavity design improves the polariton injection efficiency by a factor of 4, as compared to a conventional microcavity design, when broad excitation pulses are centered at a wavelength of $\lambda = 740$ nm. Most remarkably, this improvement factor reaches 270 when the excitation wavelength is centered at 750 nm.

## I. INTRODUCTION

Recent technological developments in the fabrication of semiconductor microcavities have allowed scientists to tackle polaritons arising from the strong coupling between the interband excitonic resonance in quantum wells (QWs) and the photonic mode of a cavity.[1] The mixed light-matter nature of these bosonic quasiparticles provides them with unique properties such as a small effective mass (5 orders of magnitude lighter than the free electron mass), allowing them to achieve dynamical Bose-Einstein condensation at elevated temperatures (between 5 and 300 K).[2-7] While such a macroscopic quantum state is reached in atomic systems by decreasing the effective temperature, in solid state systems, one must increase the overall density of polaritons $\rho$ to enable an efficient energy relaxation channel through excitonic interactions. In general, polaritons are injected inside the microstructure by optical pumping. At a particular excitation fluence, corresponding to a threshold polariton density $\rho_{th}$, the system enters a regime of stimulated bosonic scattering characterized by a fast increase of the energy ground state occupation. The resulting macroscopic state is coherent and it is referred to as a polariton condensate. Recently, there has been a growing interest to combine, in a single experiment, such a solid state condensate and ultrashort pulses, where the latter can serve both as a time-resolved diagnostic tool and an activation stimulus.[8-12] Notably, such a scheme was used to image the inner dynamics of a solid- state condensate and reveal fundamental differences between the regimes of polariton lasing and photon lasing.[8] However, the injection of polaritons with spectrally broad femtosecond pulses can be quite inefficient since conventional microcavity structures, characterized by a broad photonic stop-band, act on the optical pump like a quasi-perfect mirror. Because higher pump fluences are then required to reach the regime of polariton condensation, these types of experiments pose strong demands on the minimum power delivered by the optical source. Additional limitations come into play if this source must be operated at wavelengths far from its optimal operation range, or when the excitation spot size must cover a relatively large sample area.[8,13,14] Under these circumstances, it becomes crucial to enhance the polariton injection efficiency. Until now, this efficiency has been considered to be intrinsically determined by the specifications of the optical source. In this letter, we use a novel approach based on an engineered sample structure. We implement a custom design of a planar GaAs/AlGaAs microcavity to optimize the polariton injection efficiency that can be achieved with a broad class of excitation laser systems: Ti:sapphire amplifiers delivering femtosecond pulses in the near-infrared region. This design can be applied to a variety of semiconductor microcavities and could pave the way to the realization of new ultrafast experiments that involve intense femtosecond pulses interacting with an optically injected polariton condensate.

---

[a]Electronic mail: jean-michel.menard@mpl.mpg.de

Conventional planar semiconductor heterostructures fabricated to investigate the regime of polariton condensation consist of an $n\lambda_o/2$ optical cavity ($n$ = 1, 2, ...) confined within multiple $\lambda_o/4$ layers of alternating semiconductor compounds, where $\lambda_o$ is the wavelength associated with the cavity mode.[2,15] These $\lambda_o/4$ layers form a set of two distributed Bragg reflectors (top and bottom) that ensure a high cavity Q factor. QWs are placed inside the cavity at the antinodes of the photonic mode to enable a strong coupling between the QW excitonic transition and the resonant cavity light field. This strong coupling leads to the formation of polaritonic states. The polaritons are typically injected in the system via non-resonant optical excitation inducing unbound electron-hole pairs in the QWs. As the free carriers lose their excess energy via scattering processes, they form excitons, i.e. hydrogen-like electron-hole pairs, which can then couple to the photonic cavity mode to form polaritons. Such a non-resonant pumping scheme is commonly used in polariton condensation experiments to prevent the coherence of the optical excitation to be imprinted on the coherence of the macroscopic quantum state.[2]

Efficient optical excitation can only be achieved when some specifications are fulfilled by the optical pump source and the microcavity sample. For instance, the excitation photon energy $\hbar\omega_P$ must be higher than the QW bandgap energy $E_g$, to achieve non-resonant excitation, but lower than the bandgap energy of all other semiconductor compounds entering in the composition of the structure, to avoid depletion of the optical pump in the distributed Bragg reflectors. Then, $\omega_P$ must be set to one of the minima in the reflectivity spectrum of the structure corresponding to a leaky mode of the microcavity. Due to the periodic nature of the sample, these minima occur at regular intervals in frequency. Finally, the pump spectral linewidth should not considerably exceed the width of one of these reflectivity dips, or else a large fraction of the excitation light is simply reflected. In this context, the choice of an efficient optical pump source is usually limited to a range of tunable continuous wave or picosecond pulsed lasers, while ultrafast systems, delivering broadband femtosecond pulses, are generally discarded. Here we demonstrate that the conventional design of a microcavity can be modified to engineer the reflectivity spectrum, while keeping all other sample parameters fixed, to enable efficient optical injection of polaritons with spectrally broad femtosecond pulses.

## II. EXPERIMENT

We investigate the impact of the microcavity design on the minimum optical pump fluence required to achieve polariton condensation by comparing two planar GaAs/AlGaAs heterostructures. One corresponds to a state-of-the-art conventional microcavity (sample A) while the other (sample B) is a microcavity with a shifted photonic stop-band to enhance the injection efficiency of polaritons as a Ti:sapphire femtosecond source delivers ultrashort pump pulses. Figure 1(a) shows the general heterostructure design of samples A and B, which are similar in many aspects. They are both composed of an AlAs $\lambda_o/2$-cavity. Furthermore, three sets of four GaAs QWs are placed at the center and the first antinodes of the optical cavity mode.

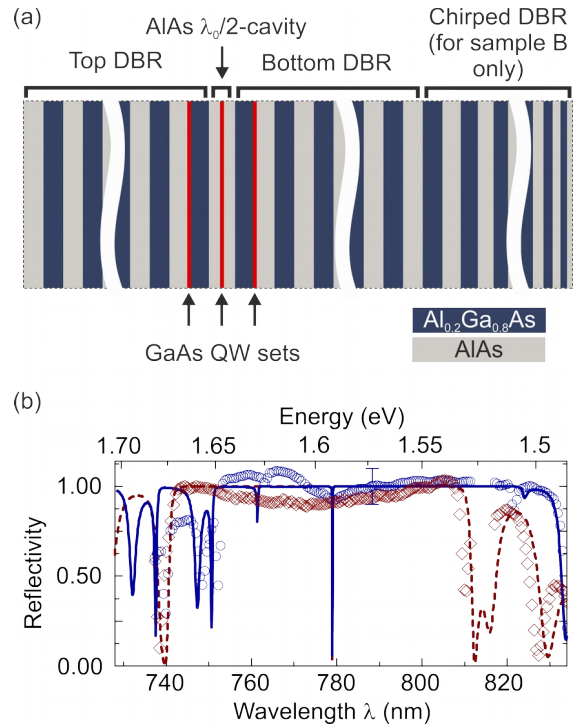

FIG. 1. (a) The general sample structure of both conventional microcavity (sample A) and pump-enhanced microcavity (sample B) contains a $\lambda_o/2$-cavity sandwiched between a top and bottom DBR. Sample B has a chirped DBR located at the back of the sample (the excitation light is incident from the left towards the top DBR). "(b) The experimental reflectivity spectra of sample A (open red diamonds) and B (open blue circles) are measured at an angle of incidence of 20° while the lattice temperature is kept at $T_L = 10$ K. The corresponding theoretical reflectivity (sample A: red dashed line, sample B: blue solid line) are obtained with the transfer matrix method.[16,17] Bulk complex refractive indices are used in the model, excitonic resonances are not included.[18,19]

The QWs are 7 nm thick and separated by 3 nm thick AlAs barrier layers. For both samples, top and bottom DBRs are formed of alternating pairs of $Al_{0.2}Ga_{0.8}As$ and AlAs layers which yield the same numerically calculated Q factor of $1.6 \times 10^4$. The samples are kept at a lattice temperature of $T_L = 10$ K and they both feature a small positive cavity-exciton detuning $\delta = 2$ meV and a comparable measured Rabi-splitting energy $\hbar\Omega_R$ of 15 meV (sample A) and 18 meV (sample B). The latter parameter determines the energy separation between the two polariton branches (lower and upper polariton) resulting from the strong coupling between the photonic mode and the excitonic resonance. Figure 1(b) shows the reflectivity spectra of both microstructures. By repeatedly shifting the central wavelength of an ultrafast Ti:sapphire optical source, we measure the broadband experimental reflectivity spectra of sample A (open red diamonds) and B (open blue circles). Each measurement is then normalized by the reflectivity spectrum obtained from a reference gold mirror inserted at the exact position of the microstructure sample. A small intrinsic tilt angle between the sample and the gold mirror however causes an experimental error in the reflectivity amplitude of ~0.10. Nonetheless, the measured reflectivity spectra show a good overall agreement with the theoretical reflectivity for sample A (red dashed line) and B (blue plain line), which are calculated with the transfer matrix method.[16,17] Most notably, we observe that the most important features of the reflectivity spectra, i.e. the measured position and width of the dips, are consistent with the theory. Major differences in their reflectivity spectra are attributed to the following disparities in the samples' structure design. First, the thickness $\lambda_t/4$ ($\lambda_b/4$) of the layers forming the top (bottom) DBR can be engineered to spectrally shift the photonic stop-band of the cavity. While sample A corresponds to a standard microcavity design defined by $\lambda_t = \lambda_b = \lambda_o$, sample B represents an asymmetrical heterostructure with $\lambda_t = 1.015\lambda_o$ and $\lambda_b = 1.030\lambda_o$, which displays additional transmission windows for non-resonant excitation at $\lambda \sim 730$ nm, 750 nm and 760 nm. Matrix transfer calculations reveal that changing the optical cavity thickness results in a spectral shift of the bare cavity mode, but does not significantly alter the position or the width of the dips in reflectivity. Such a variation of the optical cavity thickness across a sample is commonly employed to obtain regions of different cavity-exciton detunings. In our experiment, the thickness of the AlAs layer located in between the top and bottom DBR is selected to yield an identical cavity mode wavelength for both samples. Furthermore, the number of layer pairs forming the DBRs of sample B must also be increased to ensure that both cavity Q factors match exactly. As a result, the top and bottom DBR, which respectively contain 26 and 28 pairs in sample A, are now formed of 28 and 34 pairs, respectively, in sample B. Finally, an additional structure innovation based on a chirped DBR, placed after the bottom DBR, is implemented on sample B to enhance non-resonant optical pump excitation. The chirped DBR, composed of 20 pairs of $Al_{0.2}Ga_{0.8}As$ and AlAs layers with thicknesses varying linearly from $0.943\lambda_o/4$ (top) to $0.924\lambda_o/4$ (bottom), by itself, forms a high-reflectivity spectral window covering the non-resonant excitation optical wavelengths between $\lambda = 710$ nm to 760 nm (Fig. 2). In a simple picture, this periodic structure reflects the excitation beam back towards the sample, allowing a second optical pass of the pump through the QWs. Matrix transfer calculations confirm that the absorption of the non-resonant pump in the QWs is enhanced by a factor of ~2 when such a chirped DBR is added at the back of the structure. Therefore, the spectral dips observed across the non-resonant excitation window for sample B (Fig. 1(b)), do not correlate to peaks in the optical transmission but to a linear absorption inside the QWs.

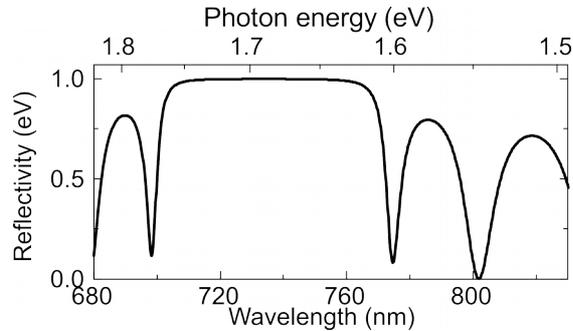

FIG. 2. Calculated reflectivity spectrum of the stand-alone chirped DBR structure obtained with the transfer matrix method for an angle of incidence of 20°.

## III. RESULTS AND DISCUSSION

Photoluminescence (PL) measurements are performed on both samples A and B. An ultrafast Ti:sapphire amplifier optical source with a repetition rate of 800 kHz is used for non-resonant injection of polaritons in the microcavities.[20] The pump beam is set at an angle of incidence of 20° relative to the microcavity's normal axis and the excitation spot has a diameter full width at half maximum (FWHM) of 35 μm on the sample surface. The PL is monitored via energy and angle resolved spectroscopy by a monochromator and CCD camera system. A linear relation holds between the emitted PL intensity and the number of cold polaritons (close to the bottom of the lower polariton branch).[2] Since both samples have similar polaritonic properties, we can draw direct comparisons between their emitted PL intensities to determine how their actual design affects injection efficiency of the polaritons at specific conditions of the optical pump. In our experiment, we investigate the PL for different pump central wavelengths $\lambda_p$ between 735 nm and 760 nm with a fixed excitation linewidth $\Delta\lambda = 10$ nm FWHM. This spectral range covers most of the purely non-resonant excitation region comprised within the lower polariton resonance and the bandgap of the $Al_{0.2}Ga_{0.8}As$ compounds ($E_g = 1.76$ eV at $T_L = 10$ K) corresponding to $\lambda_g = 705$ nm.[19]

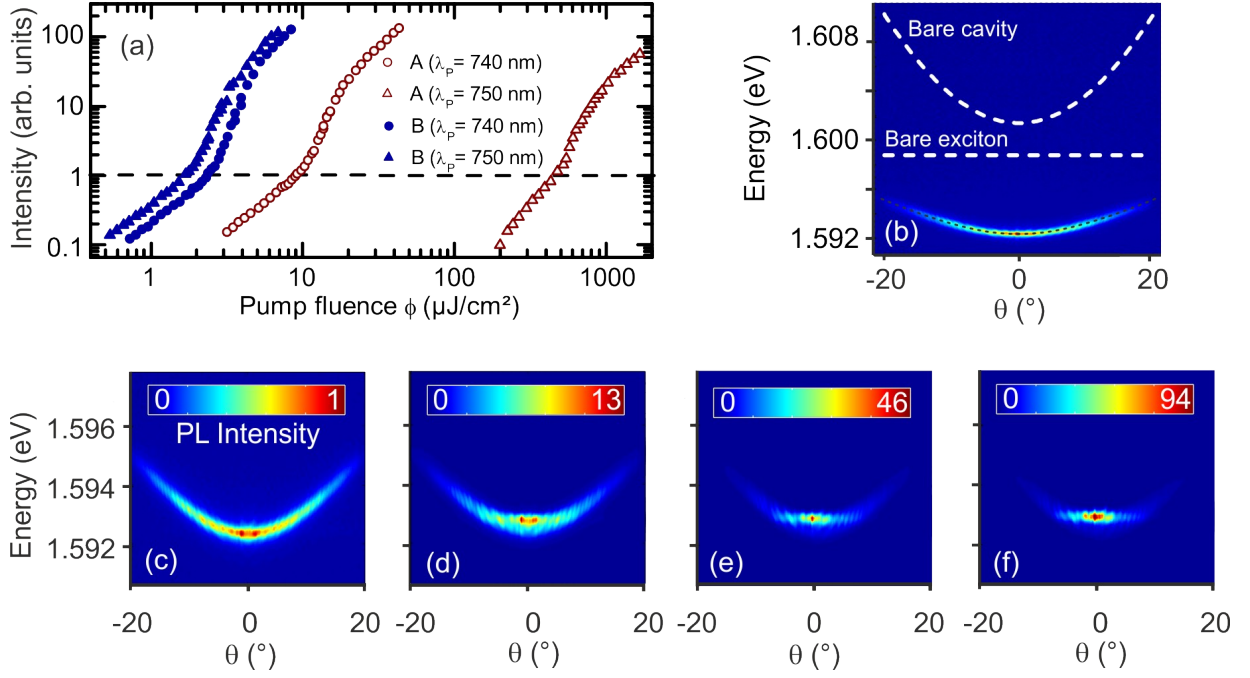

FIG. 3. (a) PL intensity measured on samples A (red open data points) and B (blue filled data points) as a function of the pump fluence $\phi$. The excitation pulses (linewidth $\Delta\lambda = 10$ nm FWHM) are centered at a wavelength of $\lambda_p = 740$ nm (circles) or $\lambda_p = 750$ nm (triangles). (b) The lower polariton dispersion (black dashed line) is fitted to the angle-resolved PL spectrum collected from sample B at a low excitation fluence and used to evaluate the dispersion relation of the bare exciton and bare cavity mode (white dashed lines).[21] (c - f) Angle-resolved PL spectra show the formation of the polariton condensate in sample B, for $\lambda_p = 750$ nm, as $\phi$ is gradually increased across $\phi_{th} = 1.7$ µJ/cm²: (c) $\phi = 0.4\ \phi_{th}$, (d) $\phi = 1.2\ \phi_{th}$, (e) $\phi = 1.4\ \phi_{th}$, (f) $\phi = 1.5\ \phi_{th}$. The sample temperature is $T_L = 10$ K.

Figure 3(a) shows the total PL intensity collected from samples A (red open data points) and B (blue filled data points) at two exemplary pump wavelengths of $\lambda_p = 740$ nm (circles) and $\lambda_p = 750$ nm (triangles). As we increase the pump fluence $\phi$, the PL intensity curves shown in Fig. 3(a), as well as all PL data monitored in our experiment, follow a threshold-like behavior. Briefly, the PL intensity follows approximatively a linear dependence with $\phi$ up to a threshold fluence $\phi_{th}$ (intersecting with the dashed line in Fig. 3(a)), above which it shows an abrupt superlinear increase. The pump fluence $\phi_{th}$ is physically associated with the critical polariton density $\rho_{th}$ corresponding to the emergence of the stimulated bosonic scattering induced by the polariton condensate.[2] This conjecture is confirmed by angle-resolved PL spectra shown in Figs. 3(b-f) for the specific case of $\lambda_p = 750$ nm. For a low excitation fluence (Fig. 3(c), $\phi = 0.4\ \phi_{th}$), light emission is spread over a broad range of angles - 20° < θ < 20°. As $\phi$ is increased, a dramatic narrowing of the angular and energetic distribution of the PL is seen (Figs. 3(d) and 3(e); $\phi = 1.2\ \phi_{th}$ and $1.4\ \phi_{th}$, respectively). Finally, at $\phi = 1.5\ \phi_{th}$ (Fig. 3(f)), most of the optical emission results from a degenerate energy state associated with the polariton condensate at θ ~ 0°. We evaluate a quantitative measure of the relative optical injection efficiency of the polaritons by extracting $1/\phi_{th}$ from the experimentally measured PL curves and their corresponding threshold fluences $\phi_{th}$. In Fig. 3(a), for instance, sample B yields a larger $1/\phi_{th}$ than sample A, by a factor of 4 at $\lambda_p = 740$ nm, and by a factor of 270 at $\lambda_p = 750$ nm. Figure 4 summarizes the complete set of PL measurements (square data points). Sample B clearly allows us to reach consistently higher values of $1/\phi_{th}$ over the full range of $\lambda_p$ investigated in our experiment. Most remarkably, our pump-enhanced microcavity design increases $1/\phi_{th}$ by more than an order of magnitude, in comparison to the conventional design, when $\lambda_p = 745$ nm, 750 nm or 760 nm. Such dramatic increase is a direct consequence of the additional dips in the reflectivity spectrum of sample B, which yields a larger integrated spectral pump absorption, as well as the presence of a chirped DBR at the back of the structure, which effectively allows a second optical pass of the excitation pulse. For both microstructure designs of sample A and B, we perform numerical calculations based on the matrix transfer method to determine the ratio between the absorbed pump power $P_{abs}$ and the incident optical power $P_{inc}$ (star shaped data points in Fig. 4). Using the experimental excitation spectrum for the calculations, we can directly compare the calculated optical power ratio absorbed in the QWs $P_{abs}/P_{inc}$, to the measured value of $1/\phi_{th}$. These data are presented in Fig. 4 and show a good agreement between theory and experiment.

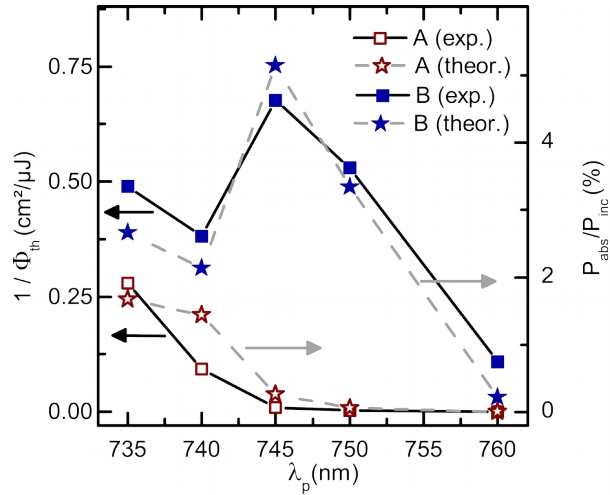

FIG. 4. Polariton injection efficiencies ($1/\phi_{th}$) determined at different central wavelengths $\lambda_P$ of the excitation pulses (linewidth $\Delta\lambda = 10$ nm FWHM) for sample A (red square data points) and B (blue square data points). The threshold fluence $\phi_{th}$ is obtained experimentally from the PL intensity measurements (similar to curves in Fig. 3(a)) and corresponds to the fluence at the injected polariton threshold density $\rho_{th}$. For both samples, the ratio between the theoretical absorbed optical power ($P_{abs}$) and the incident optical power ($P_{inc}$) (star shaped data points) is calculated with a transfer matrix method for each excitation spectrum.

The enhanced polariton injection efficiencies obtained with sample B over a spectral window extending from $\lambda_p = 735$ nm to 760 nm confirm that a sample design based on a shifted photonic stop-band can lead to lower values of $\phi_{th}$ even when the excitation pulses have a linewidth $\Delta\lambda > 10$ nm. Moreover, the fact that optical pumping can now be achieved efficiently at longer wavelengths approaching the cavity mode resonance provides two additional technical advantages. First, an ultrafast Ti:sapphire pump laser source can now operate closer to its optimal wavelength range, centered at $\lambda = 800$ nm,[22] to deliver higher output powers. Also, the local rise in the lattice temperature induced by optical pumping can now be reduced since free carriers are injected with less excess energies in the QWs.

## IV. CONCLUSION

In summary, we present a pump-enhanced microcavity sample design based on a shifted photonic stop-band that enables higher polariton injection efficiencies and a lower condensation fluence threshold over a broad range of excitation wavelengths. Non-resonant pumping of a solid state condensate can now be achieved with spectrally broad femtosecond pulses at reduced pump intensities. We demonstrate the concept for a planar GaAs/AlGaAs structure, but our design could potentially be implemented for any type of semiconductor microcavities to gain optical access over a spectral window located nearby the photonic mode frequency. We believe that our pump-enhanced microcavity design will enable future experiments to combine polariton condensates with ultrashort laser pulses providing access to new combinations of nonlinear processes.

## ACKNOWLEDGEMENTS

We thank D. Bougeard for helpful discussions, I. Gronwald and M. Furthmeier for experimental assistance. Support by the European Research Council via ERC grant 305003 (QUANTUMsubCYCLE), the German Research Foundation (DFG) via the Emmy Noether Program (HU1598/1-1), the Alexander von Humboldt Foundation, the French RENATECH network and the Labex NanoSaclay is acknowledged.